\documentclass[fleqn,10pt]{wlscirep}
\usepackage[utf8]{inputenc}
\usepackage[T1]{fontenc}
\title{Sorting of mesoporous silica derivatives by random optical fields}
\author[1]{Mohammad Hadi Sadri}
\author[1]{Ramin Jamali}
\author[2,3]{Asif Jamal Khan} 
\author[3,4]{Fozia Rehman}
\author[1,5,*]{Ali-Reza Moradi}
\affil[1]{Department of Physics, Institute for Advanced Studies in Basic Sciences (IASBS), Zanjan 45137-66731, Iran}
\affil[2]{Shaanxi Key Laboratory of Earth Surface System and Environmental Carrying Capacity, College of Urban and Environmental Sciences, Northwest University, Xi'an, China}
\affil[3]{S$\tilde{\rm{a}}$o Paulo State University (UNESP), Institute of Chemistry, Araraquara SP, Brazil}
\affil[4]{Interdisciplinary Research center in Biomedical Materials (IRCBM), COMSATS University Islamabad, Lahore Campus, Pakistan}
\affil[5]{School of Nano Science, Institute for Research in Fundamental Sciences (IPM), Tehran 19395-5531, Iran}

\affil[*]{moradika@iasbs.ac.ir}

\begin{abstract}
Mesoporous silica particles are promising candidates for drug delivery applications. In this paper, we first synthesize mesoporous silica MCM-41 and its derivative MCM-41GA  with anchored glutaraldehyde bridges, and characterize them using a variety of techniques, including nitrogen adsorption/desorption, X-ray diffraction, NMR spectroscopy, scanning electron microscopy, and thermogravimetric analysis. 
Then,  we employ random optical fields to sort mesoporous silica particles. 
Random optical fields by containing  local intensity gradients throughout a wide range of field of view provide an elegant, easy-to-implement, and  low-cost  variant of multiple optical tweezers, which is known as  speckle  tweezers (ST). ST, similar to  multiple optical tweezers, for  manipulation tasks, such as  trapping, sorting, and guiding of collection of micro and sub-micro objects in several disciplines including statistical physics, chemistry, microfluidics and material science. 	We show that ST can restrict, sieve, and sort MCM-41 and MCM-41GA particles. The different interaction of  mesoporous silica variations with the applied  ST may be attributed to the pre-applied modification and the differences in the porosity structure and distribution. Therefore, the results provide insight into the textural and chemical characteristics of mesoporous materials, contributing to a deeper understanding of their potential applications.
\end{abstract}
\begin{document}

\flushbottom
\maketitle
%
%
\thispagestyle{empty}


	\section{Introduction}
\noindent
Mesoporous silica represents a highly efficient class of inorganic adsorbents and was primarily investigated because of its unique properties, which include high specific surface areas (up to $ \rm{1500 \; m^{2}g^{-1}} $), large pore volume (up to $ 1.5 \; {\rm{cm^{3}g^{-1}}} $), uniform pore size, stable inorganic skeleton, well-ordered structure, and hydrothermal stability \cite{wu2013synthesis,tomasello2013chikungunya,rehman2018organo}. 
The shape and size of  mesoporous silica  strongly depend on the surfactant used for the synthesis of these materials, and  there are many types of mesoporous silica. Amongst, MCM-41 is the most studied material \cite{khan2020mesoporous}.  
MCM-41 type mesoporous silica with hexagonally ordered cylindrical pores was first reported in 1992 by Mobil Corporation Scientists, using cationic surfactants under basic conditions \cite{khan2020mesoporous,talavera2019effect,grunberg2004hydrogen,coasne2006adsorption}. The aforementioned physiochemical properties along with a range of organic functional groups such as amines, carboxylic acid, sulfonic acid, and thiols render MCM-41 as an excellent candidate for a variety of applications like drug delivery, dye adsorption, ion separation, sensing, catalysis \cite{khan2020mesoporous,xu2018iron,ying2018nanocoating,wei2018preparation}.

In this research, we use a light-matter interaction process called speckle tweezers (ST), which incorporates random optical fields to sort mesoporous silica materials MCM-41 and the mesoporous silica materials  that have undergone modifications  inside a microfluidics chip. 
Study of light-matter interaction has led to invention of several novel  techniques and has deeply impacted many research areas. 
One of the important breakthrough techniques that revolutionized measurement and applying very small forces and torques at microscopic and nanoscopic scales, has been the invention of Arthur Ashkin, namely, optical tweezers or optical trapping (OT) \cite{PhysRevLett.24.156,Ashkin:86,jones2015optical}. 
Since its invention, OT has been effectively implemented in two scales: the sub-nanoscale for cooling atoms, ions and molecules \cite{jones2015optical,PhysRevLett.68.3861,RevModPhys.85.1083} and the micro-scale to non-invasively manipulation of microscopic objects \cite{nuemanOT}. Indeed, awarding the Nobel prizes in 1997 and  2018 to these two developments reflects the highest impact of OT.  
OT at micro-scale has been extensively used in several disciplines including biology and medicine, statistical physics, nanotechnology and chemistry  \cite{jones2015optical,nuemanOT,jonassort,onofrionano}. 

In the simplest case, OT is performed through focusing a laser beam with a Gaussian intensity profile by a high NA microscope objective.
The basic principle of OT is the existence of intensity gradient at the focus of the laser beam which provides the trapping force to confine dielectric particles. 
On the other hand, a number of essential researches and applications in biomedicine and microfluidics often require controlled manipulation of collection of micro- and nano-scale particles \cite{biomedicine,biomedecine2,grier2003revolution}. 
To this end, multiple optical traps, which can be created by proper engineering the laser beam wavefront through the use of spatial light modulators \cite{Multitrap}, or by time sharing of the beam between multiple trap sites through beam steering devices, such as acousto-optic deflectors \cite{emiliani2004multi}. In these approaches the arrays of trap sites are  pre-designed and usually created in regular grids.
It is obvious that for several applications, in which manipulation of several microscopic particles is the required task,  a regular array of trap site is not required, on the one hand, and the particles do not require to be tightly confined, on the other hand. 

In recent years, speckle tweezers (ST) have attracted significant attention in the field of optical trapping due to their flexibility in  manipulation, compact structure, and easy and low-cost implementation. 
A speckle field is an intensity distribution of  laser light that is formed by coherent superposition of  light waves  with random phase or amplitude. Speckle fields can be generated by various processes, such as back-scattering  of laser light from a rough surface, passing of laser light through  complex  media,  or mode-mixing in a multimode optical fiber \cite{Goodman}. 
At first glance, speckle  seems to be a disturbing phenomenon to be suppressed as, indeed, it is in imaging and interferometry systems \cite{deSP,deSP2,yamaguchi2006phase}.  However, speckle formation  may be  considered  as an advantageous   phenomenon,  because it contains valuable statistical information about the sample or process that generates the pattern. This positive feature of speckle pattern has led to the growing field of laser speckle analysis and comes along with numerous applications in life science, industry and technology \cite{OtherS,OtherS2,nazari2023laser,panahi2022detection,pedram2023evaluation,jamali2023surface,rad2020non}. 
Additionally, due to that speckle patterns contain bright regions surrounded by a connected network of dark sites,  there are lots of local intensity gradients throughout the pattern, each can be considered as an optical trap site. Therefore a speckle pattern can be used as an irregular array of optical traps. This feature of the speckle patterns is called speckle tweezers (ST), which may be used to trap, guide, sieve and sort plenty of micro particles simultaneously according to different kind of their physical properties  \cite{STVolpe,Stepbystep,STimpo}. 
One of the first experiments on the use of ST  demonstrated the emergence of anomalous diffusion in colloids and showed the control of the motion of Brownian particles \cite{giygan}. Also, by converting the high-intensity speckle grains into the corresponding thermal speckle grains the motions of multiple particles on plasmonic substrates were controlled \cite{Opto}. 
In another study, it has been shown that the density of the structural defects in a 2D binary colloidal crystal can be engineered using a speckle field \cite{2Dbinary}. A memory equation based on a theoretical model was described the motion of colloidal particles in the presence of ST field \cite{STimpo,memory}. As explained above,  the speckle field contains a random, but somehow tunable distribution of isolated high-intensity regions surrounded by irregular and interconnected low-intensity regions. On the other hand,  low refractive index particles are rippled by positive intensity gradients and need to be trapped, for example, by Laguerre-Gaussian beams which contain central dark regions. Given this,    ST has been used for manipulation of several high and low refractive index particles simultaneously \cite{jamali2021speckle}.

The paper is organized as the following. In Section \ref{section2} we provide the sample preparation and the ST experimental procedure. In Section \ref{section3} the characterization results are explained and ST results are presented and discussed. The paper is concluded in Section \ref{section4}.

\section{Material and methods}
\label{section2}
\subsection{Synthesis of mesoporous silica MCM-41}
Analytical grade solvents/reagents were purchased and used as received. Triethyl orthosilicate (TEOS) (Sigma Aldrich), structure-directing Cetyltrimethylammonium bromide (CTAB), 3-aminopropytriethoxysilane (APTES), glutaraldehyde (GA) were Sigma Aldrich products. Ethanol, methanol, ammonium hydroxide, and hydrochloric Acid (HCl) are Synth products. Deionized water was used in the entire work.  

Mesoporous silica MCM-41 was synthesized using an established procedure \cite{khan2020mesoporous}. In a $500 \;$mL conical flask, $7.28 \;$g of CTAB was dissolved in a mixture of deionized water ($259.2 \;$mL), ethanol ($32.4 \;$mL), and $20.9 \;$mL  aqueous ammonia was added to the reaction container after full dissolution. Then $22.3 \;$mL  of TEOS was slowly added while vigorously stirring. The above solution had the following molar composition: $0.1$ TEOS: $0.02$ CTAB: $2.4 $ $NH_{4}OH$: $5.2$ $C_{2}H_{5}OH$: $14.4$ $H_{2}O$. The synthesized silica MCM-41 with CTAB was washed with $200 \;$mL deionized water after $4$ hours of stirring and dried for $3$ hours in air at $373 $K. 
The Schiff's base method was used to design surface-modified mesoporous silica MCM-41GA. In this procedure, 3-aminopropyltriethoxisylane was reacted in a $2:1$ ratio with glutaraldehyde (GA) using a protocol developed by our group\cite{rehman2014applicability}. For this purpose, about $8.4 \;$mL ($36.0 \;$mmol) of APTES is reacted with $1.7 \;$mL ($18.0 \;$mmol) of GA in $50.0 \;$mL of ethanol using $1.0 \;$mL triethylamine as a catalyst. This mixture was stirred for $72$ hours at $50^{\circ}$C  under a dry nitrogen atmosphere before moving to a three-necked circular bottom flask containing $1.0$ grams of MCM-41 in xylene. The resultant product was allowed to stir for additional three days at $ 75^{\circ}$C under vacuum to yield MCM-41GA.

\begin{figure}[tb!]
	\centering
	\includegraphics[width=.7\textwidth]{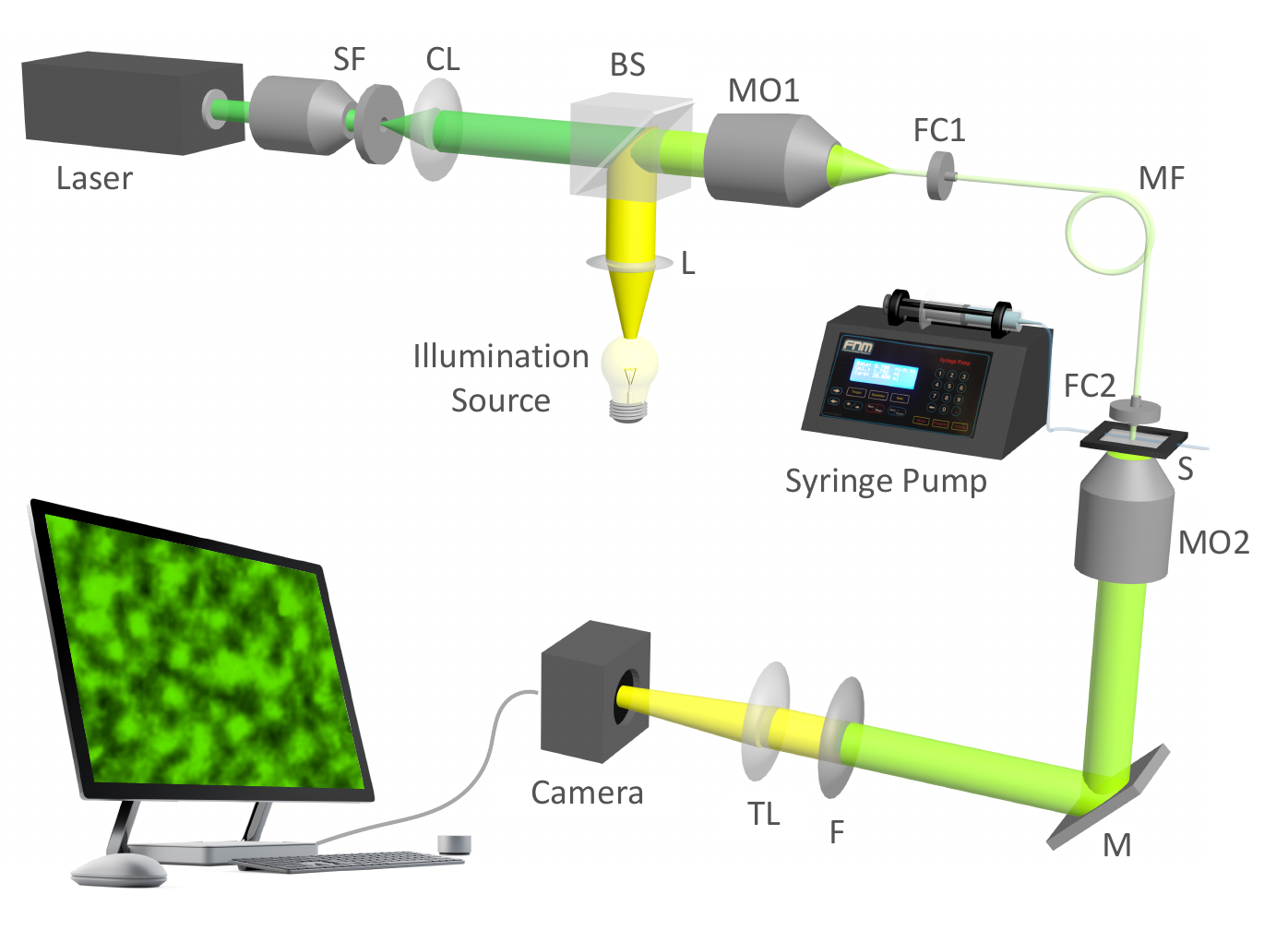}
	\caption{Experimental setup to generate random optical fields via laser propagation through a multimode fiber. SF: spatial filter, CL: collimating lens, TL: tube lens, L: lens, BS: beam-splitter, MO: microscope objective, FC: fiber coupler, MF: multimode fiber, S: sample, M: mirror, F: filter.}
	\label{fig1}
\end{figure}
\subsection{ST experimental setup to generate random optical fields}
The ST setup for random optical field generation and its application on the present material is schematically shown in Fig. \ref{fig1}.
In this configuration, a solid-state laser (Shafa Parto Parse Co., PSU-FC, 532 nm,  1 W) is used as a coherent light source. 
The laser beam is spatially filtered by the use of spatial filter (SF) which consists of  an optical microscope objective and a pinhole. The diverging clean beam is collimated by  the collimation lens (CL) and is sent onto a focusing objective (MO$1$, Nikon, 10$\times$, NA = 0.25). The focused laser beam is coupled to a multi-mode fiber (MF) by placing a coupling mount (FC1) right in the focus of MO$1$. The core diameter of MF is  365 $\pm$ 14 $\mu$m, and the refractive indices of the core and the cladding  are 1.4589 and 1.4422, respectively. The effective numerical aperture of the MF is 0.22. 
Random mixing of spatial modes propagating through the MF  leads to the irregular interference pattern, i.e., the speckle pattern at the output of the MF (mounted on FC2) and shines the sample (S). The length of the optical fiber is $L_{\rm{MF}}=80 \;$cm, and it is carefully fixed on mounts to avoid geometry caused changes on the output light field.  The maximum attenuation of laser power by the MF may be calculated by $A= \frac{10}{L_{\rm{MF}}} \log _{10}\frac{P_{in}}{P_{out}}$, where $P_{in}$ and $P_{out}$ are the input and output powers, and, given the above maximum laser power, is $\cong 15.5  \frac{dB}{Km}$. 
The  speckle pattern generation by MF have significant advantages over  rough surface back-scattering or diffusive medium transmission methods. The nature of an optical fiber causes sufficient flexibility and portability, and more importantly by proper adjustment of sample-to-fiber distance, demanded  grain sizes of the speckle field may be obtained. Nevertheless, the statistics of the grain sizes throughout the speckle field follows a normal distribution \cite{Goodman,bender2018customizing}. The probability density function of the speckle pattern intensity, however,  follows the negative exponential (gamma with parameter 1) distribution, i.e., $\frac{1}{\langle I\rangle} \exp(-\frac{I}{\langle I\rangle})$ \cite{Goodman,bender2018customizing}.

The sample is subjected to bright field microscopic imaging  using a microscope objective (MO$2$, 10$\times$, NA = 0.25), mirror (M), tube lens (TL) and the camera (DCC1645C, Thorlabs, 8 bit dynamic range,  5.2 $\mu$m pixel pitch). A beam-splitter (BS) is used for  sending  the  white light source  in a common path with the laser light for proper microscopic imaging of the sample. Lens (L) collects the highly diverging illumination light.  In order to track the particles trajectories reliably in the follow-up image processing stage, we also filter (F) the trapping laser light out by the use of an interferometric filter which rejects the 532 nm laser light. 
In order to demonstrate the capability of the ST method for  mesoporous  particle sorting,  MCM-41 and  MCM-41GA particles are infused  into the sample chamber by the use of a syringe pump (FNM, SP102HSM) at 0.5 $\frac{\rm{mL}}{\rm{hr}}$ flowing rate. The sample chamber is a  rectangular-shaped capillary tube (VitroCom, path length 400 $\mu$m, width 5 mm, length 10 cm). 
The output speckled light of the MF illuminates the channel and affects the local motions of the particles when passing through the speckle light field. The whole process is live monitored, recorded and stored as sequences of  images, to be post-processed.

\section{Results and discussion}
\label{section3}
The samples that are used for speckle-based sorting experiments are characterized by several techniques, and the results are discussed in the following. 

\subsection{Nuclear magnetic resonance (NMR) spectroscopy}
\begin{figure}[tb!]
	\centering
	\includegraphics[width=0.5\textwidth]{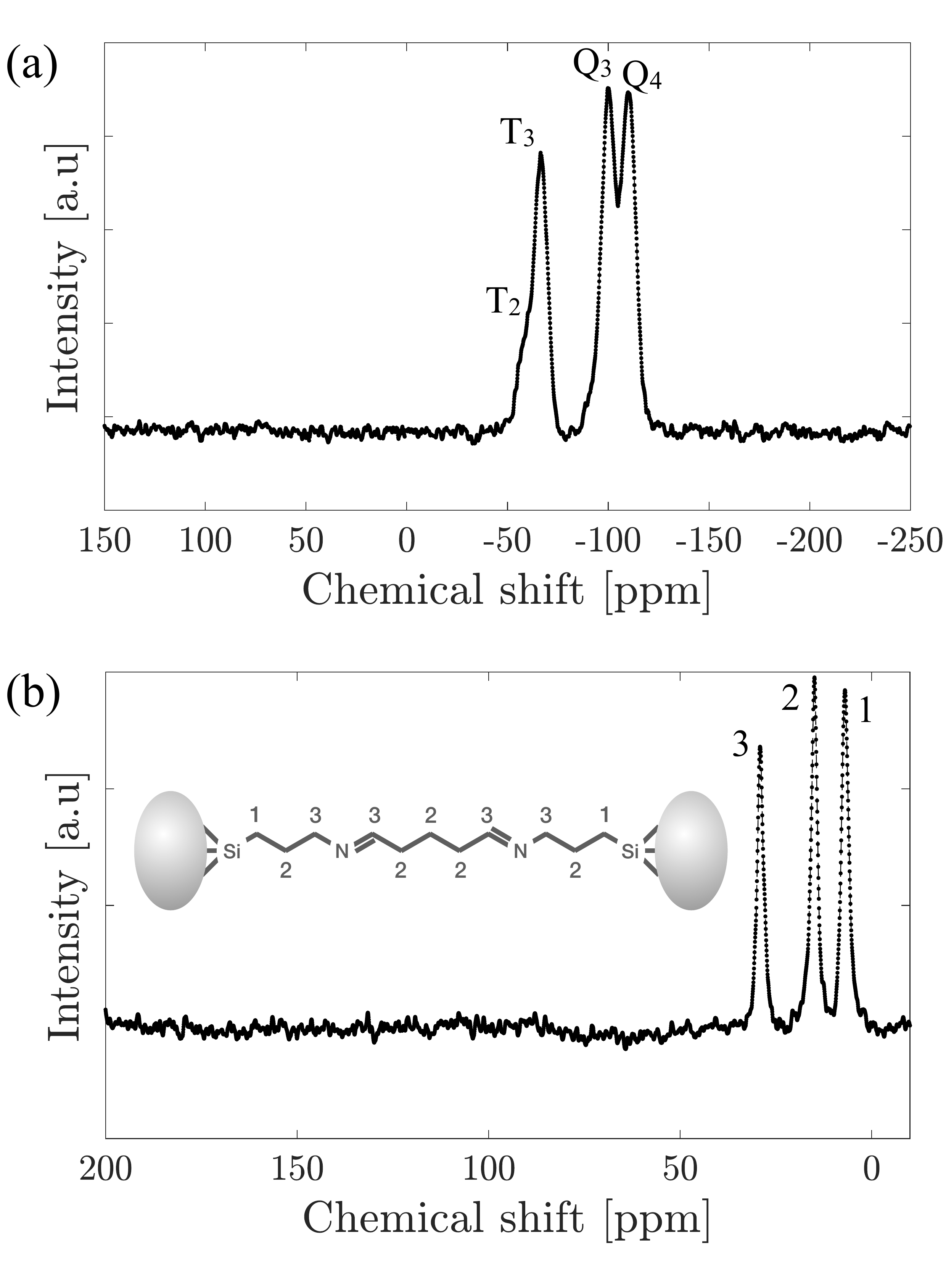}
	\caption{(a) $^{29}$Si CP/MAS NMR spectra of silica MCM-41GA. (b) $^{13}$C NMR spectra of modified silica MCM-41GA. }
	\label{fig2}
\end{figure}

The NMR spectra of the solid samples are obtained at room temperature using a Bruker Avance $3-300$ MHz spectrometer. 
For each run, approximately one gram of solid sample is compacted in $4$ mm zirconium oxide rotors. With a spinning magic-angle of $10$ MHz, measurements of silicon and carbon nuclei are taken at frequencies of $59.63$ MHz and $75.47$ MHz, respectively. $^{26}$Si  and $^{13}$C  CP/MAS spectra are obtained with pulse repetitions of $3\;$s for both nuclei and contact times of $4\; $ms and are shown in Fig. \ref{fig2}.

Surface modification of MCM-41 with GA bridges is confirmed with $^{29}\rm{Si}$ CP/MAS NMR spectroscopy (Fig. \ref{fig2}(a)). The presence of Q4 $ Si  (OSi)_{4} $, Q3 $[ (OSi)_{3} (OH)]$, T3 $ [RSi (OSi)_{3} ] $, and T2 $[RSi (OSi)_{2} (OH) ] $ (R= organic group) signal in spectrum of MCM-41GA suggests that the prepared bridges are attached to the silica surface in a bi- or tridentate manner. The greater strength of T signals indicates that the prepared organic chains have been effectively substituted the surface silanol groups. 

The spectra of the $^{13}C$ nucleus for MCM-41GA, in Fig. \ref{fig2}(b), on the other hand, shows  various signals. The signals at 9.0; 21.0; 42 are attributed  to C-Si, C-C, and C-N bonds. The chemical shift appears at 167.3 ppm for C=N \cite{rehman2014applicability}. A proposed structure of MCM-41GA is shown in the inset of  Fig. \ref{fig2}(b) \cite{khan2020mesoporous}. The appearance of these chemical shifts indicates that the surface modification with the synthesized bridge has been successful.

\subsection{N$_{2}$  Adsorption/desorption and elemental analysis}
Nitrogen $N_{2}$ adsorption/desorption (BET) is a non-destructive method to analyze mesoporous and microporous materials. 
Micromeritics ASAP $2000$ is used to adsorb and desorb nitrogen in the relative pressure range of $0.06$ to $0.99$ in $0.015$ increments at T=77 \;K. Samples are degassed for $8$ hours at 363 \;K.  
The Barrett-Joyner-Halenda (BJH) procedure is used to measure the pore diameters and their distributions from the desorption branches of the experimental isotherms \cite{zhao1998nonionic}. The pore volumes are measured at a relative pressure of $0.99$. 
The pore size of MCM-41GA decreases by $1.7$ nm compared to its original material MCM-41 with a pore size of $2.0$ nm. BET surface area for MCM-41GA is measured as $449.0 \;  \rm{m}^{2} /\rm{g}$. The surface area of Mesoporous silica significantly decreases after surface modification due to the attachment of organic groups with the active silanol groups.

Quantitative elemental analyses are performed using a Perkin-Elmer PE-$2400$ system. 
The organic content attached to the MCM-41GA surface is estimated with elemental analysis and is given in Table\ref{tab01}. The C and N contents are estimated as $7.42$ and 2.5 \; mmol {g}$^{-1}$, respectively, which indicates that all active silanol groups of MCM-41 react with synthesized organic bridges. The predicted and measured carbon to nitrogen molar ratio ($\frac{C}{N}$) is close to each other. These results show that the MCM-41 silica modification process using synthesized monomers succeeded effectively.  

\begin{table}[]
	\centering
	\caption{Percentages (\%) of carbon (C) and nitrogen (N) and their respective amounts in mmol g$^{-1}$, the expected and the calculated of carbon/nitrogen molar ratios for MCM-41 and MCM-41GA.}
	\label{tab01}
	\begin{tabular}{|c|c|c|c|c|c|c|}
		\hline
		Sample     & C [\%] & N [\%] & C {[}mmol g$^{-1}${]} & N {[}mmol g$^{-1}${]} & $\frac{C}{N}$ (Exp.) & $\frac{C}{N}$ (Calc.) \\ \hline\hline
		MCM - 41   & 0.3  & 0.03 & 0.25                                     & 0.3                                      & -                    & -                     \\ \hline
		MCM - 41GA & 8.9  & 2.5  & 7.42                                     & 25                                       & 5                    & 4                     \\ \hline
	\end{tabular}
\end{table}

\subsection{Scanning electron microscopy (SEM)}
\begin{figure}[tb!]
	\centering
	\includegraphics[width=.7\textwidth]{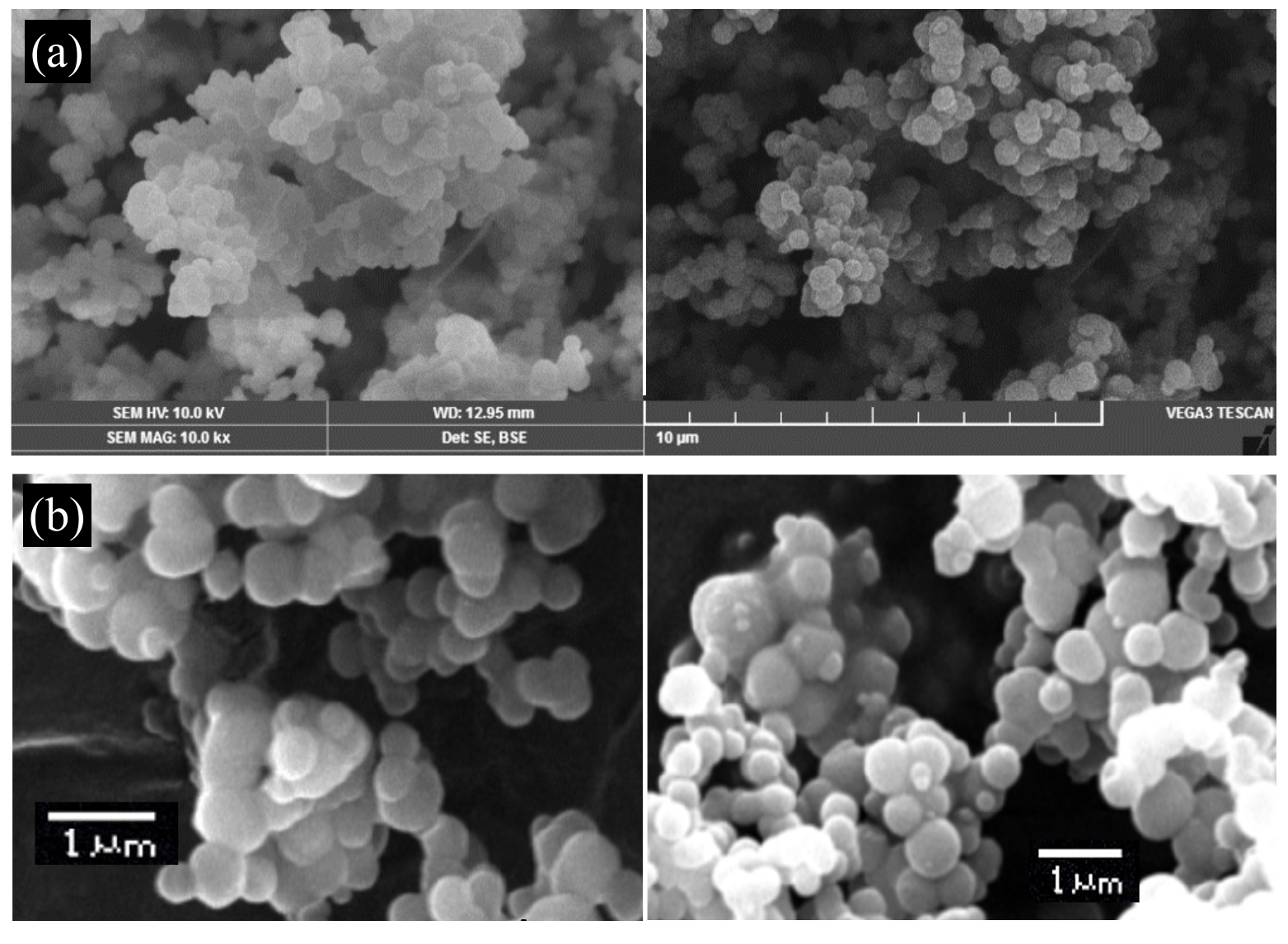}
	\caption{SEM images of (a) MCM-41 and (b) MCM-41GA.}
	\label{fig3}
\end{figure}

SEM micrographs of the prepared mesoporous silica samples are taken with a JEOL JSM-6360 LV scanning electron microscope operating at 20 kV (Fig. \ref{fig3}). Samples are mounted on C tape with a Bal-Tec MD20 kit before the study.
The particle shape is spherical, with an average particle size of about $0.5 \;  \mu$m. For  MCM-41GA, the SEM micrograph shows that the original shape and size remain preserved after modification with GA bridges. The shape and size of this material remain the same as those of MCM-41. The particle size is in the range between $0.4$ to $0.6\;  \mu$m, and the average particle size is measured as $0.65 \;  \mu$m \cite{hobson1997bridged,he2011mesoporous,khan2020mesoporous}.
No noticeable change in size and shape is observed, which means that the modification process does not affect the original architecture of MCM-41.

\subsection{X-Ray diffraction (XRD)}
\begin{figure}[tb]
	\centering
	\includegraphics[width=.8\textwidth]{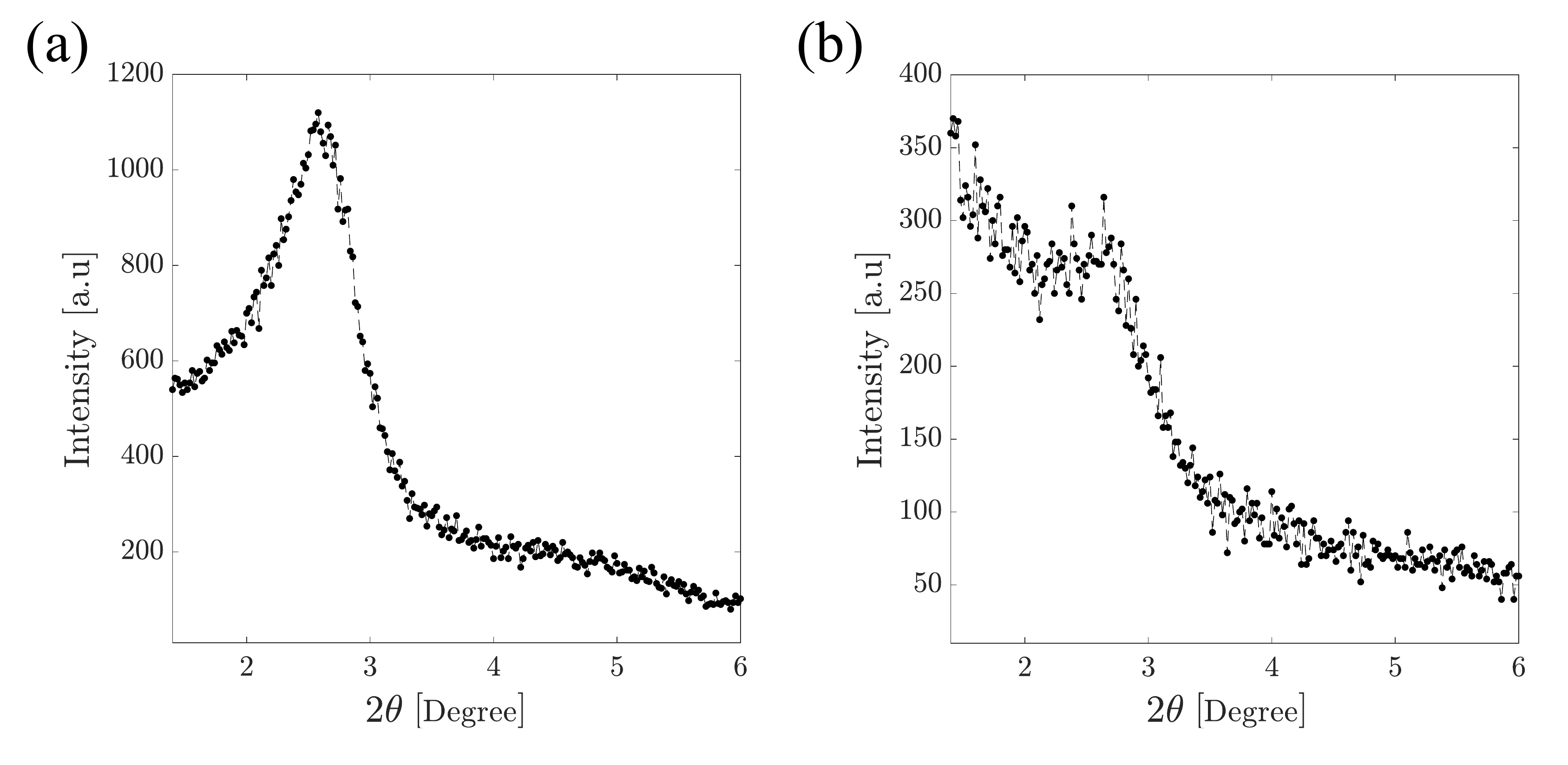}
	\caption{X-ray diffraction of (a) MCM-41 and (b) MCM-41GA. }
	\label{fig4}
\end{figure}

XRD and small-angle X-rays diffractograms are obtained using X-rays scanner model XDR 7000 with a scanning rate of $ 2^{ \circ } \; \frac{2 \theta }{\rm min} $ with Cu K-$\alpha $ radiation source using $ \lambda =0.154$ nm. 
XRD patterns of modified silica MCM-41 and MCM-41GA are shown in Figs. \ref{fig4}(a) and (b), respectively. These materials show a significant diffraction peak at (100) and two minor peaks at (110) and (200). However, these peaks' intensities reflect the silica surface's alteration. These diffraction patterns are the characteristic peaks MCM-41, as reported in \cite{khan2020mesoporous}, showing ordered 2D meso-structured with hexagonal p6mm symmetry\cite{dai2014amino}. 

\subsection{Thermogravimetry}
\begin{figure}[tb!]
	\centering
	\includegraphics[width=.7\textwidth]{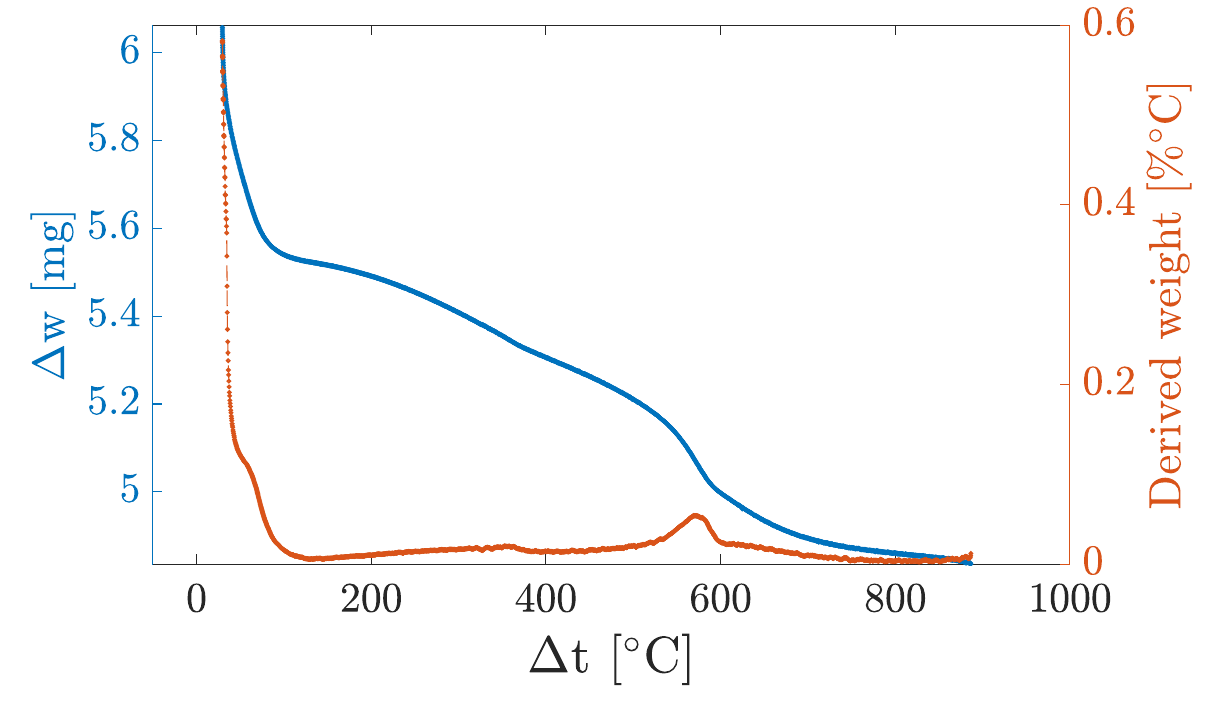}
	\caption{Thermogravimetric and derivative curves of modified silica  MCM-41GA. }
	\label{fig5}
\end{figure}

Thermogravimetric curves are obtained using a TA tool and a model 1090 B thermobalance in an argon atmosphere with a flow rate of 30 $\frac{\rm mL}{\rm s}$.
Thermogravimetric analysis of modified materials and the derivative curves are shown in Fig. \ref{fig5}. Mass loss is observed over a different temperature range. The highest mass loss is observed for modified material MCM-41GA in the temperature range of 200 to 600 $^{\circ}$C. This shift is proportional to the sum of organic groups connected to the silica network's inorganic backbone. The initial mass loss of around $ 100^{ \circ}$C  is due to water and the $ \Delta m $ for the last stage ($800-1000 ^{\circ}$C) can be attributed to silanols condensation to form siloxanes and water.

\subsection{Manipulation with random optical field and tracking particles}

The MCM-41 and  MCM-41GA particles under ST are live monitored and their bright field microscopic images are stored as  sequences of image files. We use ImageJ software for  image processing and MATLAB software for presenting the data. 
The ST-sample interaction can change the  motion of particles, and therefore, the position tracking of an ensemble of particles will provide information about the trajectory of the samples and, therefore, their  behavior when interact with the speckled light. 
As explained in the Introduction the  ST applies local intensity gradients which are the reasons of optical local forces. 
The mechanism of ST as an extension for optical trapping is explained in detail in \cite{jamali2021speckle}. 
The effect of ST is assessed visually by the microscopic image sequences and then in a more quantitative fashion  by trajectory analysis and particle velocimetry. 

\begin{figure}[t!]
	\centering
	\includegraphics[width=\textwidth]{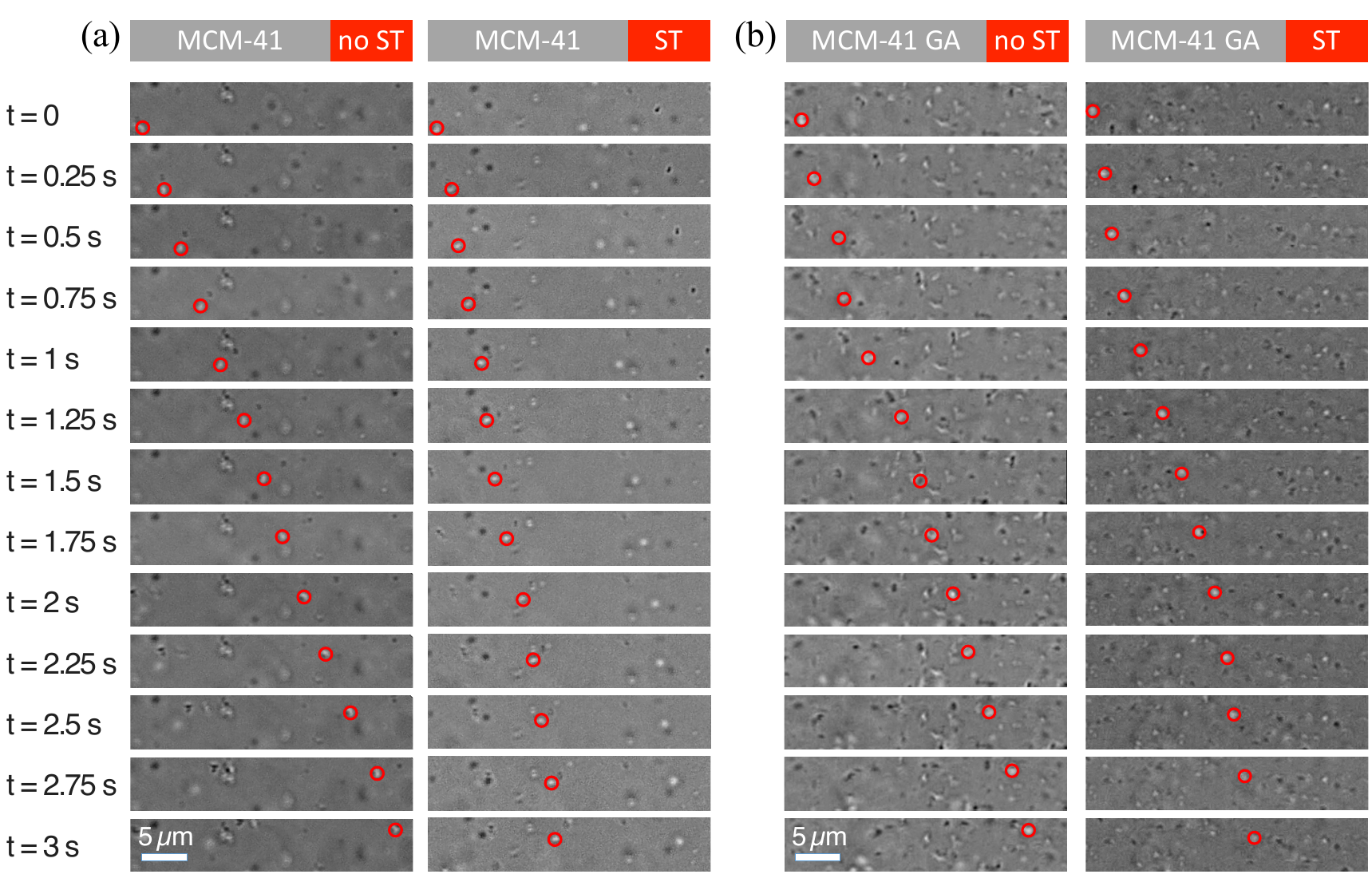}
	\caption{(a) MCM-41 and (b) MCM-41GA particles when (left) the laser is switched off and (right) under the illumination with a random laser  field.}
	\label{fig6}
\end{figure}

Figures \ref{fig6}(a) and (b) show the microscopic images  of MCM-41 and  MCM-41GA particles, respectively, taken during the infusion to the sample chamber.  The microscopic images are acquired for 3 s at 25 fps, but only the snaps associated to  every 0.25 s are shown. In each panel the left and right columns show the particles when the laser light is switched off and when the  speckle pattern illuminates the sample, respectively. A single particle (red circled) is chosen  in each case and followed in time.  For both MCM-41 and MCM-41GA cases it is obvious that ST hinders the particles' speed. In the absence of laser light, in 3 s,  the indicated MCM-41  particle travels about  25 $\mu$m and the  MCM-41GA particle travels about 22 $\mu$m.  However, when the sample is under ST the indicated MCM-41 particle travels  only about 12 $\mu$m, i.e. it travels much slower. Moreover, the indicated MCM-41GA, while  travels slower than the no-ST case, but its speed is hindered less than the MCM-41 case when shone by ST light. The behavior is also visible  for other particles which are not red circled.  The demonstration of Fig. \ref{fig6} concludes that ST affects both  mesoporous silica particles, but the influence is different for the modified derivatives. This, indeed, is the key to sort them in a very simple, easy-to-implement, non-invasive and cost-effective way, i.e., simply flowing the mixture through a fluidic channel and shine them with a random laser light.  

\begin{figure}[t!]
	\centering
	\includegraphics[width=.8\textwidth]{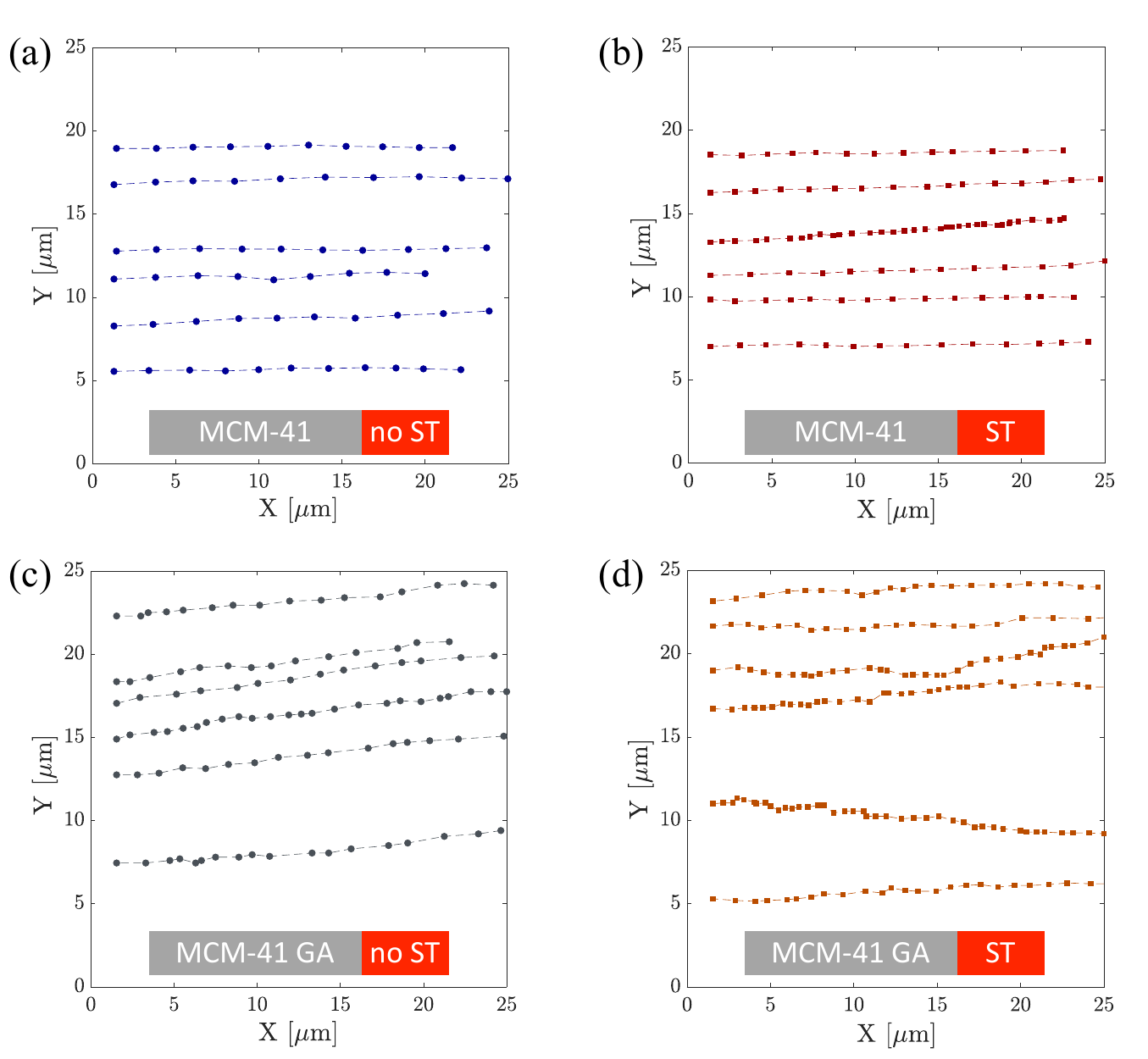}
	\caption{Trajectories of  MCM-41 (a) when the laser is switched off and (b) under the illumination with a random laser  field, and (c-d) similarly for MCM-41GA particles. Each data point is associated with a single acquired frame. }
	\label{fig7}
\end{figure}

\begin{figure}[t!]
	\centering 
	\includegraphics[width=0.6\textwidth]{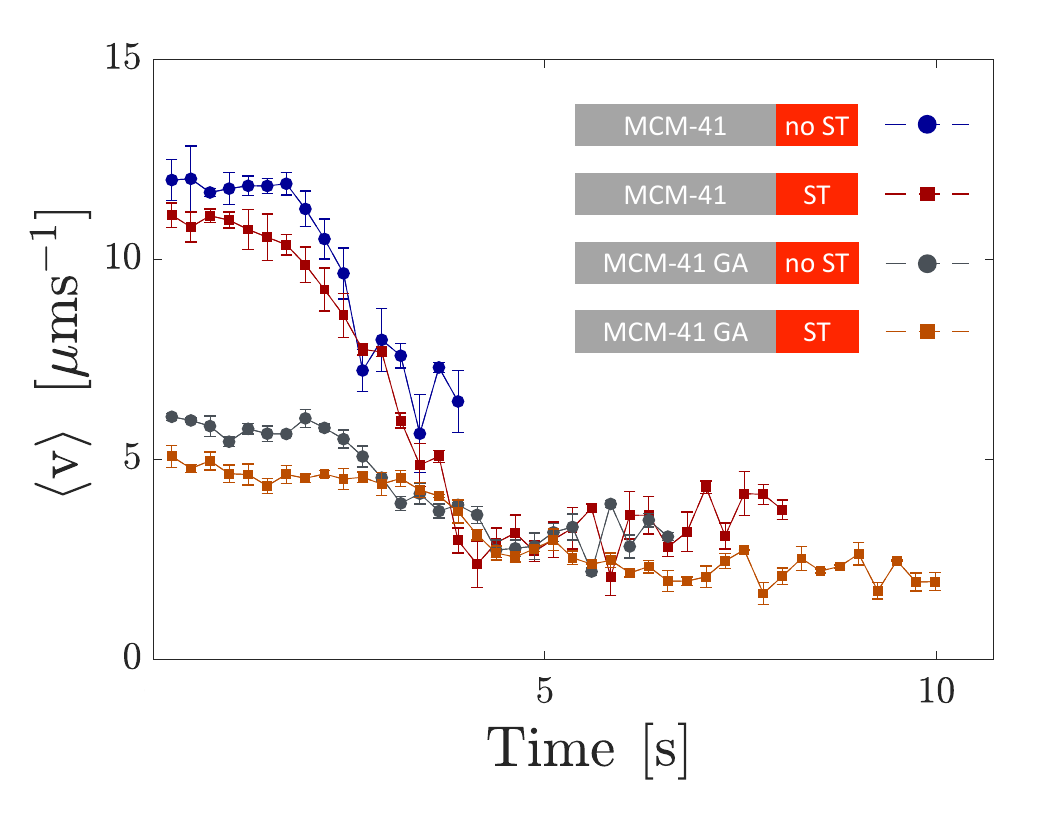}
	\caption{Average velocity over  ensembles of MCM-41 and MCM-41GA particles when the laser is switched off and when  illuminated with a random laser field. A larger number of data points in no-ST cases is due to the faster travel of particles in the field of view of the acquisition device. }
	\label{fig8}
\end{figure}

The above  procedure is repeated several times for each case, and the same behavior is observed. However, we assess the results in a more quantitative fashion. Figure \ref{fig7} shows the trajectories of multiple particles in each case, in which, given the identical acquisition frame rate, the number of data points for a specific travel distance is different. This verifies that the aforementioned concluding behavior applies for all the particles in the field.  In Fig. \ref{fig8} in each two successive frames of microscopic images we  calculate the velocities of several (at least an ensemble of 40) particles and report their average versus time. The time interval between consecutive data points is 0.25 s. 
Apart the appearance of some fluctuations in the values of $\langle \rm v\rangle$, it is obvious that the velocities in the presence of ST and the optical forces are hindered for both types of the particles. However, the effect is more pronounced for the MCM-41 particles. This difference may be  attributed to the different levels of hydrophilicity in MCM-41 and MCM-41GA. The presence of silanol groups allows water and other polar molecules adsorption inside the mesoporous of MCM-41, giving a hydrophilic character to the surface. Surface modification with Glutaraldehyde via crosslinking amine silylating agent leads to change in surface charges and zeta potential as previously reported \cite{rehman2014applicability} and eventually changes the hydrophilic character of original precursor silica MCM-41.
Change of in the particle velocity of MCM-41GA in comparison to MCM-41 in the presence of light field can be attributed to change in physiochemical characteristics of MCM-41 after modification process, in which surface free silanols (-Si-OH) of MCM-41 surface are reacted with the synthesized GA-bridged long chains (Fig. \ref{fig2}(b)).  Both hydrophobic character and surface area are responsible for the changed behavior of MCM-41GA, when compared to parent material MCM-41.

 
\section{Conclusion}
\label{section4}
In conclusion, we introduce speckle tweezers on fluidic chips as a simple yet useful non-contact, non-invasive, and  flexible approach for sorting mesoporous silica particles with different modifications, which are important  candidates as drug delivery carriers.  
We prepared mesoporous silica MCM-41 and a version modified with glutaraldehyde bridges, MCM-41GA, and   characterized them thoroughly using nitrogen adsorption/desorption, X-ray diffraction, NMR spectroscopy, scanning electron microscopy, and thermogravimetric analysis methods. 
Then, we showed that the presence of speckle tweezers applies different hindering in the flowing velocity of the two mesoporous particles. The experimental results, graphically and quantitatively, verifies the effect of random light fields on the behavior of microscopic particles. The different interaction of mesoporous silica variations with the applied random light field  may be attributed to the pre-applied modification and the differences in the porosity structure and distribution. 
The presence of silanol groups allows water and other polar molecules adsorption inside the mesoporous of MCM-41, giving a hydrophilic character to the surface. Surface modification with Glutaraldehyde via crosslinking amine silylating agent leads to change in surface charges and zeta potential  and eventually changes the hydrophilic character of original precursor silica MCM-41. Both hydrophobic character and surface area are responsible for the changed behavior of MCM-41GA, when compared to parent material MCM-41.
   The platform can serve for different targeted and collective control of particles based on their size, weigh, modification and structure, which indeed covers a variety of different samples. Therefore,  the approach may be considered for various applications, such as microfluidic manipulation, and particle-based diagnostics for optical manipulation tasks especially for drug delivery structures. 

\bibliography{MSE2024MCM_refs}

\section*{Acknowledgment}
\noindent
The authors would like to thank  Maria  Sotomayor and Mojtaba Khorasani for the fruitful discussions and Zaib us Sama for his assistance in sample preparation. F. Rehman acknowledges  the support of CAPES-PRINT - PROGRAMA INSTITUCIONAL DE INTERNACIONALIZACAO 41/2017 and INCT-DATREM, Brazil.
\section*{Competing interests}
The authors declare no competing interests.

\section*{Data availability}
The datasets used or analyzed during the current study available from the corresponding author on reasonable request.

\end{document}